\documentclass[conference]{basi}

\usepackage[british]{babel}
\usepackage{rotating}
\usepackage{dcolumn}

\usepackage{rotating}
\usepackage{dcolumn}
\begin{document}
\title[High-redshift galaxies]{Stellar Spectral Signatures in High-Redshift Galaxies}
\author[C. Leitherer]%
       {Claus Leitherer\thanks{email: \texttt{leitherer@stsci.edu}}\\
       Space Telescope Science Institute, 3700 San Martin Drive, Baltimore, MD 21218, USA}

\pubyear{2014}
\volume{10}
\pagerange{\pageref{firstpage}--\pageref{lastpage}}


\maketitle
\label{firstpage}

\begin{abstract}
Stellar emission and absorption lines are routinely observed in galaxies at redshifts up to 
$\sim$5 with spectrographs on 8 -- 10-m class telescopes. While the overall spectra are well 
understood and have been successfully modeled using empirical and theoretical libraries, some
challenges remain. Three issues are discussed: determining abundances using stellar
and interstellar spectral lines, understanding the origin of the strong, stellar
He II $\lambda$1640 line, and gauging the influence of stellar Ly-$\alpha$ on the 
combined stellar+nebular profile. All three issues can be tackled with recently
created theoretical stellar libraries for hot stars which take into account
the radiation-hydrodynamics of stellar winds.
\end{abstract}


\section{Introduction}
High-quality rest-frame ultraviolet (UV) spectra of large numbers of galaxies at
redshifts up to $\sim$5 are routinely obtained with current 8 -- 10-m class telescopes
(e.g., Jones et al. 2012). Gravitational lensing and the associated magnification factors allow pushing
the detection limit and achievable signal-to-noise to the extreme limit (see Bayliss et
al. 2013 and Stark et al.
2013 for detailed studies and surveys of lensed galaxies, respectively). The age of the universe at $z=4$ is about 
1.6~Gyr; therefore the observed galaxies are young and currently star-forming. This
makes the rest-frame UV where the bulk of the luminosity is emitted the wavelength
region of choice. In the context of this conference, which focuses on stellar
spectral libraries, I will address several issues related to the stellar aspects
of the observed rest-frame UV spectra: (i) the potential of stellar lines to
determine abundances as an alternative to emission-line techniques; (ii) evidence
of variations of the initial mass function (IMF) with $z$; (iii) the properties 
of stellar Ly-$\alpha$ as determined by wind effects.

\begin{figure}
\centerline{\includegraphics[width=9cm]{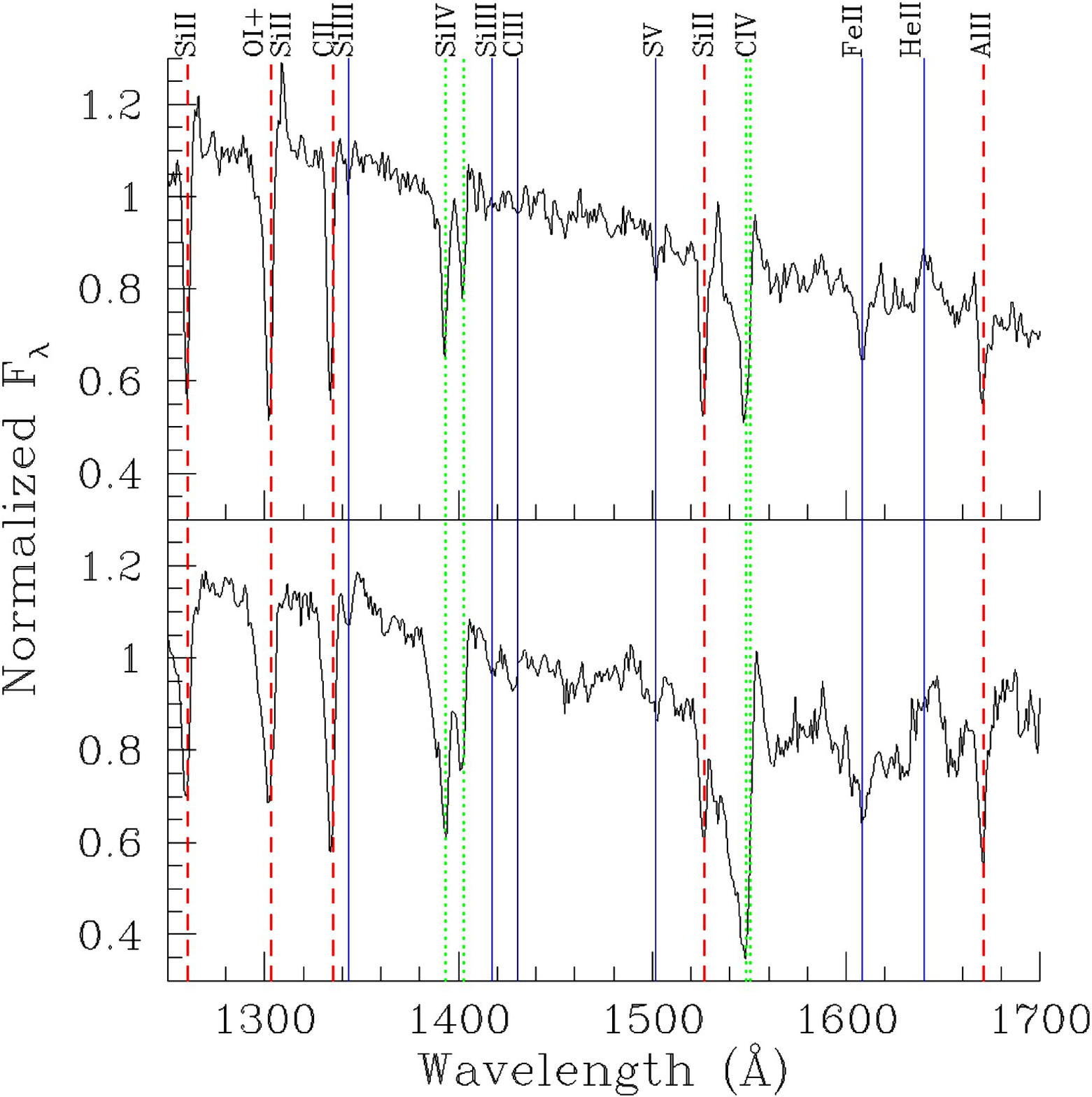}}
\caption{Comparison of the average rest-frame UV spectrum of 811 star-forming galaxies (Shapley et al. 2003; top) and a stacked UV spectrum of 16 local star-forming galaxies (bottom).
Solid vertical lines indicate stellar photospheric lines, dotted lines are for stellar-wind lines, and dashed lines denote interstellar lines. From Schwartz et al. (2006).}
\end{figure}

\section{Abundance determinations from UV spectra}
Traditional abundance determinations in star-forming galaxies rely on emission-line techniques.
At high $z$, the observed wavelengths of many of the most widely used rest-frame optical nebular lines
are shifted into the near-infrared (IR). Despite advances in sensitivity, obtaining high signal-to-noise
spectra in the near- or mid-IR is still more challenging than in the optical. Moreover,
atmospheric transparency is small between the J, H, K, and L windows so that often not all 
the required lines can be measured simultaneously. The UV spectra provide an alternative means
of determining abundances.

In Fig. 1 a comparison of the UV spectral region of high- and low-$z$ galaxies is shown. Note
the close correspondence of the two spectra, suggesting that locally calibrated spectral
libraries should do a good job in interpreting distant galaxies as well. Three classes
of lines can be identified in this figure: (i) broad, shallow photospheric blends such as 
C~III $\lambda$1425, (ii) broad P Cygni-type stellar wind lines such as C~IV $\lambda$1550,
and (iii) narrow interstellar absorptions such as C~II $\lambda$1335.

\begin{figure}
\centerline{\includegraphics[width=5cm,angle=-90]{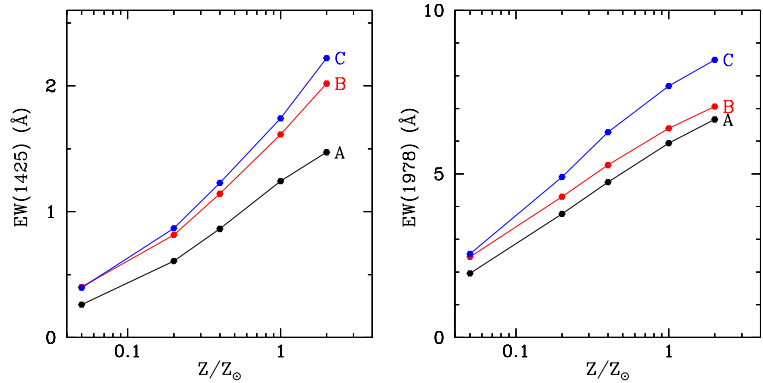}}
\caption{Equivalent widths of the C~III $\lambda$1425 (left) and Fe~III $\lambda$1978 (right)
blends versus chemical composition. The three curves in each plot are for IMFs with
a Salpeter (A) or Miller-Scalo (B) slope, and a Salpeter slope truncated at 30~M$_\odot$ (C).
From Rix et al. (2004).}
\end{figure}

Rix et al. (2004) created a theoretical stellar library of OB stars for the Starburst99 code 
(Leitherer et al. 1999; Leitherer \& Chen 2009) and predicted the strengths of several
photospheric blends as a function of chemical composition. An example is in Fig.~2, which
shows the results for the (mostly) C~III and Fe III blends. The features are only mildly sensitive 
to IMF variations since both lines and continuum arise from the same stellar mass range. Using
metallicity sensitive photospheric lines for abundance determinations is analogous to classical
methods that are widely used for early-type galaxies.
The Rix et al. models were expanded by Leitherer et al. (2010) who included the strong
stellar-wind lines in the spectral library. Despite being optically thick, these lines are
metallicity indicators since they depend on the mass-loss rates, which in turn depend
on the metal abundance because the winds are driven by metal lines. Note that these lines
are quite sensitive to the IMF as well. James et al. (2013) performed a comparison of the
various metallicity indicators with the gravitationally lensed galaxy CASSOWARY20 at $z = 1.4$.
The continuous spectral coverage from 1300 to 9000~\AA\ (rest-frame) provided by the VLT X-shooter
spectrograph permitted a comparison of both the UV and optical (nebular) techniques. Good
agreement was found. 

Heckman et al. (1998) and Leitherer et al. (2011) suggested yet another empirical metallicity indicator: interstellar
absorption lines. The strengths of these lines (e.g., C~II $\lambda$1335 correlate with chemical
composition despite being heavily saturated. The likely interpretation is a correlation 
between metallicity and starburst strength and eventually increased macroscopic interstellar
turbulence due to energy input by winds and supernovae.
Ultimately, abundances derived from UV spectra of high-$z$ galaxies
should provide invaluable constraints on the cosmic chemical evolution of galaxies, which
is a major quest in contemporary observational cosmology (Yuan et al. 2013).

\section{The origin of the He II $\lambda$1640 line}

The behavior of the major stellar lines in the spectra of high-$z$ galaxies is well understood
and can be reproduced with current spectral libraries. A notable exception is the 
He~II $\lambda$1640 line, which is weak in local star-forming galaxies but can become the
strongest stellar feature in UV rest-frame spectra at high $z$ (Shapley et al. 2003; Erb et al.
2010). The line has a width of order $\sim$1000~km~s$^{-1}$, which excludes in interstellar 
origin. Stars capable of producing observable He II must meet one or more of the following:
He overabundance, high surface temperature, and/or dense stellar winds. These conditions
are most naturally met in Wolf-Rayet stars, which are related to very massive, evolved O stars
(Crowther 2007). The stronger He II at high $z$ could indicate a larger fraction of very
massive stars compared to the local universe, and therefore a flatter IMF. A flattening
of the IMF with $z$ is not unexpected: for instance, cosmological simulations by Dav\'e (2008)
can account for the observed trend of the star-formation rate versus mass relation with $z$ by
invoking an excess of massive stars at high redshift. 

The currently implemented UV spectral library in Starburst99 (Leitherer et al. 2010) does not include high-resolution
spectra of Wolf-Rayet stars. In order to investigate the nature of the bona fide Wolf-Rayet features
observed at high $z$, we expanded our library by including Wolf-Rayet spectra. We constructed
a library of Wolf-Rayet spectra from the grid of Potsdam Wolf-Rayet 
models\footnote{http://www.astro.physik.uni-potsdam.de/\~{}wrh/PoWR/powrgrid1.html}.   
The spectra are calculated from model atmospheres which account for non-LTE, spherical expansion and 
metal-line blanketing (Gr\"afener et al. 2002; Hamann \& Gr\"afener 2003; 2004). We selected
12 models each for the nitrogen and the carbon sequence, covering temperatures between 30,000~K
and 160,000~K. Details will be published in a forthcoming paper (Leitherer et al., in
preparation).  

In Fig.~3 the spectra of an evolving single stellar population are shown at ages between 1 and 7~Myr.
Around 3~Myr, strong He~II $\lambda$1640 appears. This line is produced by the Wolf-Rayet stars
in the new spectral library. As expected, the line is very sensitive to IMF variations and
can be used to probe an excess or deficit of very massive stars. As a cautionary note, the
line strength is critically dependent on the Wolf-Rayet star parameters, which are predicted 
by stellar evolution models. The evolution models are quite uncertain in this respect and
may in fact disagree with observations (Brinchmann et al. 2008; Eldridge \& Stanway 2009).

\begin{figure}
\centerline{\includegraphics[width=9cm,angle=90]{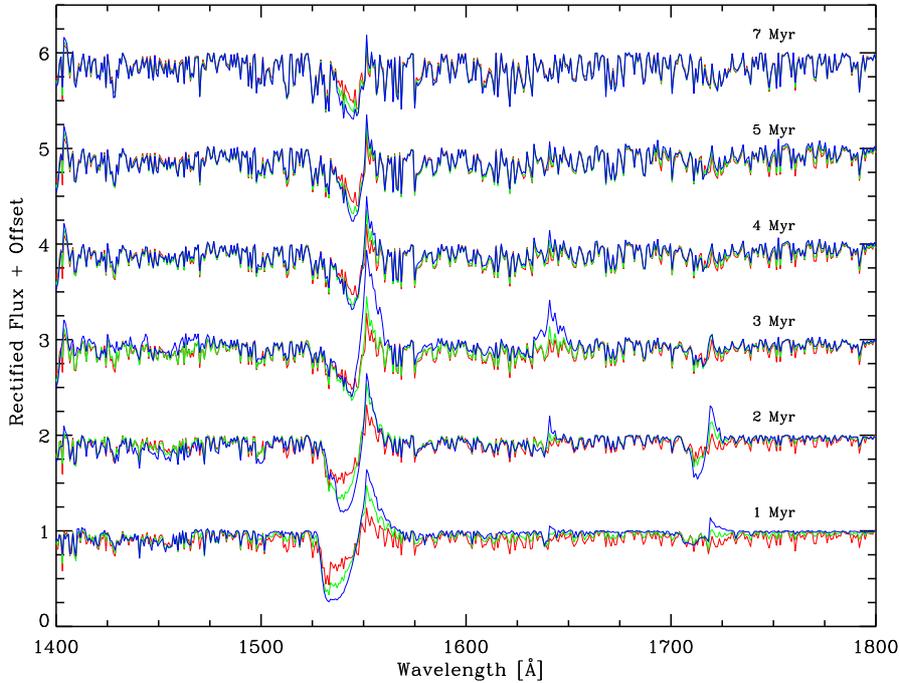}}
\caption{Evolution of a single stellar population between 1 and 7 Myr. The most
conspicuous features are C~IV $\lambda$1550 (at any time step), He II $\lambda$1640
(between 2 and 4~Myr), and N~IV $\lambda$1720 (between 1 and 5~Myr). The three
spectra at each epoch are for IMF slopes 3.3 (red), 2.3 (green), 1.0 (blue), with flatter
(= smaller exponents) slopes producing stronger lines. Note the sensitivity of the
He II line to the IMF.}
\end{figure}

\section{Ly-$\alpha$ in OB stars}
Pe\~na-Guerrero \& Leitherer (2013) used CMFGEN (Hillier \& Miller 1998) and TLUSTY (Lanz \&
Hubeny 2003; 2007) blanketed non-LTE atmospheres
for compiling a grid of stellar Ly-$\alpha$ equivalent widths in OB stars. The stellar
Ly-$\alpha$ line in star-forming galaxies is not of primary interest per se but it is for its
diluting effect on the total nebular+stellar profile. Nebular Ly-$\alpha$ emission is a 
prime star-formation tracer at high $z$ (Dijkstra \& Wyithe 2012), yet the correlation between star-formation rate 
and Ly-$\alpha$ luminosity is much more complex than expected purely for a recombination
line. While dust plays a major role, the effect of underlying stellar absorption on
the net profile is under debate (Charlot \& Fall 1993; Valls-Gabaud 1993; Schaerer \& 
Verhamme 2008). 

\begin{figure}
\centerline{\includegraphics[width=8cm]{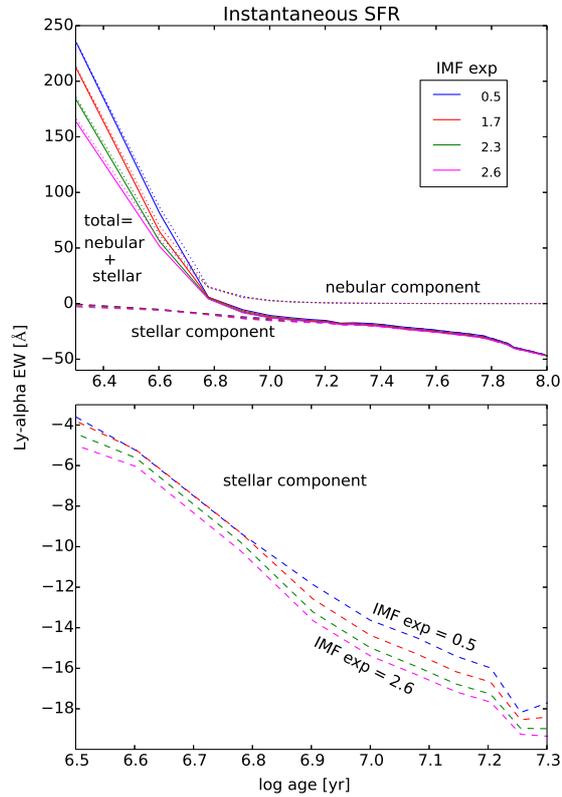}}

\caption{Evolution of stellar and interstellar Ly-$\alpha$ equivalent width in a single stellar population. Upper:
stellar, nebular, and total equivalent width for different IMF exponents; lower: zoomed view of the stellar 
component. From Pe\~na-Guerrero \& Leitherer (2013).}

\end{figure}

Intrinsic Ly-$\alpha$ is not observable even in the closest O stars due to dominating
interstellar H~I absorption. Consequently we must rely on model atmospheres. The success of
the models in predicting other, observable lines gives confidence in the theoretical approach.
Typical equivalent width values are close to zero for the hottest and most luminous stars, whose
stellar winds produce P Cygni profiles with comparable emission and absorption components. B stars
have Ly-$\alpha$ in absorption with equivalent values of tens of \AA.

In Fig. 4, the equivalent values of Ly-$\alpha$ for a single stellar population predicted by Starburst99 are
plotted. Two distinct regimes can be identified: for ages younger than $\sim$10~Myr, nebular Ly-$\alpha$ 
is large and stellar Ly-$\alpha$ is small, and the stellar line in negligible. Conversely, after 10~Myr,
the supply of ionizing photons quickly decreases so that the strong B-star Ly-$\alpha$ absorption
takes over. Incidentally, for typical starburst ages of 10~Myr, the stellar and nebular contributions
are comparable.

\section{The future}

Most UV libraries of hot stars currently in existence are theoretical. While model atmospheres for
hot stars are more trustworthy than their cool counterparts, observational comparisons are nevertheless
crucial. Such tests have successfully been done out to the Magellanic Clouds (Massey
et al. 2013). However, limiting the comparison to stars in the Magellanic Clouds imposes severe
restrictions. Arguably the most crucial unknown is the influence of chemical composition, and 
truly metal-poor massive stars, whose lifetimes are only a few tens of Myr, do not exist in our Local Group
of galaxies.

In order to enter a new parameter regime, stars in galaxies with oxygen abundances of a few percent 
the solar value must be studied. Such galaxies exist at a distance around 10 to 20 Mpc, the most famous being
I~Zw~18 at 18~Mpc. I~Zw~18 is $370 \times$ more distant than 30~Doradus, and 1~arcsec corresponds to 88.5~pc. 
A 20-m class telescope will have a spatial resolution of 0.1~pc at 1500~\AA. A typical O star in 
I~Zw~18 has a flux level of $\sim 10^{-18}$~erg~s$^{-1}$~cm$^{-2}$~\AA$^{-1}$ at 1500~\AA, which is 
higher than the sky background. In terms of photometry, a B0~V star has $V \approx 27$, and a G2~V star has $V \approx 36$.
Such observations are outside the capabilities of existing facilities but within the reach of the 
Advanced Technology Large Aperture Space Telescope (ATLAST). ATLAST is a NASA strategic mission concept study for the next generation of UVOIR space 
observatory\footnote{http://www.stsci.edu/institute/atlast}.
With a primary mirror diameter of close to $\sim$20 m it has a sensitivity limit that is up to 
$2000 \times$ better than that of the Hubble Space Telescope. ATLAST would be able to obtain
truly transformative observations of hot, massive extragalactic stars.

\section*{Acknowledgements}

Wolf-Rainer Hamann and Helge Todt (Potsdam) provided invaluable
help with the download and generation of the Wolf-Rayet library.

\end{document}